\documentclass[
pra,superscriptaddress,showpacs]{revtex4}
\usepackage{amsbsy}
\usepackage{amsmath}
\usepackage{amsfonts}
\usepackage{graphicx}
\usepackage{color}
\usepackage{epsfig}
\usepackage{bm}
\usepackage{amssymb}
\usepackage{amsthm}
\usepackage{multirow}
\usepackage{mathrsfs}

\newtheorem{proposition}{Proposition}

\newcommand{\ext}{ps} 

\newcommand{\rot}{270}   

\errorstopmode

\begin{document}

\title{No approximate complex fermion coherent states}
\author{Tom\'a\v s Tyc}
\affiliation{Institute of Theoretical Physics, Masaryk University,
        61137 Brno, Czech Republic}
\author{Brett Hamilton}
\affiliation{Institute for Quantum Information Science, University of
        Calgary, Alberta T2N 1N4, Canada}
\author{Barry C.~Sanders}
\affiliation{Institute for Quantum Information Science, University of
        Calgary, Alberta T2N 1N4, Canada}
\affiliation{Australian Centre of Excellence for Quantum Computer Technology,
    Macquarie University, Sydney, New South Wales 2109, Australia}
\author{William D. Oliver}
\affiliation{MIT Lincoln Laboratory, 244 Wood Street, Lexington, MA 02420, USA}

\date{January 3, 2006}

\begin{abstract}

Whereas boson coherent states with complex parametrization provide an elegant,
and intuitive representation, there is no counterpart for fermions using complex
parametrization. However, a complex parametrization provides a valuable way
to describe amplitude and phase of a coherent beam. Thus we pose the 
question of whether a fermionic beam can be described, even approximately,
by a complex-parametrized coherent state and define, 
in a natural way, approximate complex-parametrized fermion 
coherent states.  Then we identify four appealing properties of boson coherent states (eigenstate of
 annihilation operator, displaced vacuum state, preservation of product states
 under linear coupling, and factorization of correlators) and show that these approximate
 complex fermion coherent states fail all four criteria. The inapplicability of complex
 parametrization supports the use of Grassman algebras as an appropriate alternative.
 
\end{abstract}
\maketitle

\section{Introduction}\label{introduction}

As fermion optics research grows, spurred for example by applications to fermionic 
quantum information processing~\cite{Sig05}, the formal question of how
to deal with fermionic coherence mathematically becomes important.
For bosons, the natural way to describe coherence and optics is in terms
of complex-parametrized coherent states, which are eigenstates of
the mode annihilation operator, and then employ the optical equivalence
theorem~\cite{Sud63,Gla63a,Gla63b,Kla66}. A similar analysis for fermions
is forbidden because the only complex-parametrized eigenstate of fermion
annihilation operators is the vacuum state, but formally a well-behaved
fermion coherent state can be introduced by parametrizing with Grassman
numbers~\cite{Mar59} rather than complex numbers,
which overcomes challenges due to anticommutativity relations~\cite{Ber66}.
These `Grassman coherent states' are eigenstates of the fermion
annihilation operators with Grassman-valued eigenvalues~\cite{Sch53}
and formally demonstrate many of the desirable coherence properties of
bosons~\cite{Cah99}.

Although the Grassmann coherent state representation provides a
beautiful analogy to the boson coherent state and its use in quantum optics
experiments~\cite{Cah99}, the disadvantage is that the Grassmann algebra and
the resulting Grassmann coherent state do not convey a clear physical picture
of coherence. Complex parameters can be understood in terms of amplitude (the 
modulus of the complex parameter) and phase (the argument of the complex parameter)
of the field. Here we explore whether complex parametrization can be employed in
some advantageous way to analyze coherence of fermionic systems. This
analysis entails defining an approximate complex-parametrized fermionic coherent state,
identifying the four chief desirable properties of coherent states, and show that 
these approximate fermionic coherent states dismally fail every single criterion,
which leads to the conclusion that complex parametrization is not useful even 
for approximations to fermion coherent states for any of the four criteria. This 
strong and negative result further reinforces the importance of using 
Grassman numbers to connect fermion and boson coherence in one formalism.

Our paper is organized as follows.  We review the number state representation
for bosons and fermions in Sec.~\ref{number_states}, and review boson coherent
states in Sec.~\ref{CSboson}.  Then, in Sec.~\ref{CSfermions}, we show why
complex fermion coherent states in a direct analogy with the boson case are
impossible and we make an attempt to introduce ``approximate fermion coherent
states''.  In Sec.~\ref{FCF}, we derive a number of properties of
normally-ordered fermion correlators and based on these properties we show that
no multi-particle fermion state can, even approximately, factorize correlation
functions.  Finally we present our conclusions in Sec.~\ref{conclusion}.

\section{Number state representation}
\label{number_states}

In this paper we denote the boson and fermion annihilation and creation
operators by $\hat a,\hat a^\dagger$ and $\hat c,\hat c^\dagger$, respectively.
The boson (fermion) operators satisfy the following commutation
(anticommutation) relations
\begin{align}\label{CRB}
  [\hat a_i,\hat a_j^\dagger]&=\delta_{ij}\openone,\quad
  [\hat a_i,\hat a_j]=0,\quad  [\hat a_i^\dagger,\hat a_j^\dagger]=0\\
  \label{CRF}
  \{\hat c_i,\hat c_j^\dagger\}&=\delta_{ij}\openone,\quad
  \{\hat c_i,\hat c_j\}=0,\quad  \{\hat c_i^\dagger,\hat c_j^\dagger\}=0
\end{align}
with $\delta_{ij}$ the Kronecker delta and $i,j$ labeling an orthonormal set of
modes.

Number states are states with a definite number of particles in a given mode.
They are eigenstates of the particle number operator ($\hat n=\hat
a^\dagger\hat a$ for bosons and $\hat n=\hat c^\dagger\hat c$ for fermions)
such that $\hat n|n\rangle=n|n\rangle$. The lowest number state, the vacuum
$|0\rangle$, has no particle, and higher number states can be generated from
$|0\rangle$ by the action of the creation operator that raises the particle
number by one.  For bosons, this action is as follows,
\begin{equation}
  \hat a^\dagger|n\rangle=\sqrt {n+1}\,|n+1\rangle, \quad  n=0,1,2,\dots
\end{equation}
while for fermions
\begin{equation}
  \hat c^\dagger|0\rangle=|1\rangle, \quad
  \hat c^\dagger|1\rangle=(\hat c^\dagger)^2|0\rangle=0
\end{equation}
and the latter formula follows from the anticommutation relations~(\ref{CRF}).
Hence, for bosons the number of particles $n$ can have any non-negative integer
value while for fermions the only allowed values are 0 and 1.  This reflects
Pauli's exclusion principle, which states that there cannot be two fermions in
the same state.  The action of the annihilation operator on number states is
the following:
\begin{align}\nonumber
  \hat a|0\rangle&=0, \quad  \hat a|n\rangle=\sqrt n\,|n-1\rangle
                   \quad (n=1,2,\dots)\\
  \hat c|0\rangle&=0, \quad  \hat c|1\rangle=|0\rangle.
\end{align}
The extension of single-mode number to multi-mode number states in
general utilizes an arbitrary, but established, ordered set of
single-mode states $| \{\mathbf{n}\} \rangle $ with ordered
occupation numbers $\{\mathbf{n}\} \equiv \{n_1,n_2,n_3,
\cdots\}$. The established ordering accommodates the potential
sign changes that occur due to the particular algebra chosen to
represent the quantum statistics of the system. In practice, this
ordering is determined through an arbitrary, but established,
ordering of the creation and annihilation operators which act on
the vacuum state $| \mathbf{0} \rangle $ (no ordering required now
within the ket), and the consequence of this ordering is
maintained through the use of the appropriate operator algebra
used when manipulating the creation and annihilation operators.

In the case of multi-mode boson fields, the commuting algebra
introduces a simplification: the operator ordering for different
modes becomes irrelevant for commuting operators. For example, the
state with one boson in mode~1 and one boson in mode~2 can be
obtained from the two-mode vacuum in two equivalent ways:
\begin{align}
\label{multimodeb1}
 |1_1 1_2\rangle&=\hat a_1^\dagger\hat a_2^\dagger|0_10_2\rangle \\
 \label{multimodeb2}
 |1_2 1_1\rangle&=\hat a_2^\dagger\hat a_1^\dagger|0_10_2\rangle \\
 \label{multimodeb3}
 |1_1 1_2\rangle&=|1_2 1_1\rangle.
\end{align}
Although one may consider the ordering in Eq.~(\ref{multimodeb1}) to be the
established ordering, there is no distinction between the resultant state in
Eq.~(\ref{multimodeb1}) and the one in Eq.~(\ref{multimodeb2}) which uses a
different ordering. This is a direct consequence of the commuting algebra used
for boson fields.  It is for this reason that one may use the number state
notation [left-hand sides of Eqs.~(\ref{multimodeb1}) and (\ref{multimodeb2})] and
the operator notation [right-hand sides of Eqs.~(\ref{multimodeb1}) and
(\ref{multimodeb2})] interchangeably with boson fields without regard for the
established ordering between different modes.

However, ordering is important for fields which do not use a commuting
algebra. For example, the state with one fermion in mode~1 and one fermion in
mode~2 can be obtained from the two-mode vacuum in two different ways:
\begin{align}
\label{multimodef1}
 |1_1 1_2\rangle&=\hat c_1^\dagger\hat c_2^\dagger|0_10_2\rangle \\
 \label{multimodef2}
 |1_2 1_1\rangle&=\hat c_2^\dagger\hat c_1^\dagger|0_10_2\rangle \\
 \label{multimodef3}
 |1_1 1_2\rangle&=-|1_2 1_1\rangle.
\end{align}
In the fermion case, the resulting states are not identical, $|1_1
1_2\rangle\neq |1_2 1_1\rangle$, because of the fermion anti-commutation
algebra $\{\hat c_1^\dagger,\hat c_2^\dagger\}=0$ (see Eq.~(\ref{CRF})). This
illustrates the point that in general an arbitrary, but established, {\em a
priori} ordering is required. From the practical point of view, it is better to
identify a multi-mode number state by a product of creation operators that
operate on an unordered vacuum state (see the right-hand sides of
Eqs.~(\ref{multimodef1}), (\ref{multimodef2})) rather than by the ordered
occupation-number notation (see the left-hand sides of
Eqs.~(\ref{multimodef1}) and (\ref{multimodef2})).

The question of ordering fermion modes explained above becomes unimportant for
mixtures of fermion number states, as we show in
Appendix~\ref{chaoticSection}. This is significant as it is mixtures of number
states, usually in the form of chaotic states, that are most often encountered
in a real physical system.  Chaotic states of bosons and fermions are discussed
in more detail in Appendix~\ref{Appendix:chaotic}.

\section{Boson coherent states}\label{CSboson}

Boson coherent states can be introduced is several ways that emphasize its
different physical or mathematical properties.  Here we consider four
interrelated ways.

\emph{Eigenstates of the annihilation operator: ---}
A common definition is that the boson coherent state is
an eigenstate of the annihilation operator,
\begin{equation}
    \hat a|\alpha\rangle=\alpha|\alpha\rangle,\quad \alpha\in\mathbb C.
\end{equation}
This yields the expansion of a coherent state in terms of number states:
\begin{equation}
    |\alpha\rangle=\text{e}^{-|\alpha|^2/2}\sum_{n=0}^\infty\frac{\alpha^n}{\sqrt{n!}}\,|n\rangle  .
\end{equation}
To define a multi-mode boson coherent state, consider the complete set of boson
modes indexed by $k$ that can be assumed to form a countable set
$k\in\mathbb{N}$. A multi-mode boson coherent state is then an eigenstate of all
annihilation operators $\hat a_k$.

\emph{Displaced vacuum state: ---}
The boson coherent state can also be defined as a displaced vacuum state
\begin{equation}
\label{displaced}
|\alpha\rangle=\hat D(\alpha)|0\rangle
              =\text{e}^{\alpha\hat a^\dagger-\alpha^*\hat a}\,|0\rangle.
\end{equation}
In other words, the boson coherent state is a member of the orbit of the vacuum
state under the action of the Heisenberg-Weyl group $D(\alpha)$, which is
parametrized over the complex field.  This definition readily extends to
multi-mode fields where $D_k(\alpha)$ is the group action for the $k^\text{th}$
mode, and the displacement operators for different modes commute with each
other.

\emph{Linear mode coupler: ---} The third option for defining boson coherent
states is connected to their behavior under the action of a linear mode
coupler.  A linear mode coupler is an important element for transforming both
boson and fermion fields and it is realized by a beam splitter for optical
fields and as a quantum point contact for confined electron gases.  Furthermore
linear loss mechanisms can be described as linear coupling between the signal
mode and the environmental modes.  For bosons, coherent states remain coherent
states under linear loss.

The boson coherent states have the following special property.  They are the
only states, when mixed on a linear mode coupler with the vacuum state, yield a
two-mode product state, which is at the same time the two-mode product coherent
state~\cite{Aha66}.  This property is important because it tells that
multi-mode coherent states remain as multi-mode coherent states under the
action of mode coupling and linear loss.

\emph{Factorization of correlators: ---} The last possibility considered here
for defining boson coherent states is related to the normally-ordered
correlators (correlation functions)~\cite{Gla63a}
\begin{equation}
  G^{(n)}(x_1,\dots,x_n,y_n,\dots,y_1)
   \equiv\langle\hat\psi^\dagger(x_1)\cdots
     \hat\psi^\dagger(x_n)\hat\psi(y_n)\cdots\hat\psi(y_1)\rangle,
\end{equation}
whose complete factorization is identified with the property of coherence.
Here $\hat\psi(x)$ is the boson annihilation operator at the point $x$, i.e.,
the sum of annihilation operators of the complete set of modes weighted by the
spatial mode functions.

A boson coherent state is then a state for which all the normally-ordered
correlators factorize, i.e., for which
\begin{equation}\label{factorization}
  G^{(n)}(x_1,\dots,x_n,y_n,\dots,y_1)=f(x_1)\cdots f(x_n)g(y_1)\cdots g(y_n).
\end{equation}
This definition is especially important as the concept of boson coherent states
was developed by Glauber~\cite{Gla63a,Tit66} to describe coherence of the
electromagnetic field, and coherent states were defined to factorize the
normally-ordered correlators in analogy to classical states of the filed.

In fact there is some ambiguity in the definition of the boson coherent state
according to this factorizability condition. With the displaced vacuum state in
Eq.~(\ref{displaced}), the phase of the coefficient for Fock state~$|n\rangle$
is $n\arg\alpha$, but the correlator factorizes for an arbitrary phase
$\varphi_n$ while the magnitudes of the coefficients have to remain the
same. Generalized coherent states that satisfy the coherence condition were
studied by Titulaer and Glauber~\cite{Tit65}.  The boson coherent
state~(\ref{displaced}) is a special case of states in this class.

\emph{Summary: ---}
Each of the four definitions given above are equivalent, which
we prove in Appendix~\ref{appen-equiv}.
In the following section we consider extending each of these
approaches to constructing coherent states to the fermion case.

For completeness, we mention another appealing feature of boson coherent
states, namely the Gaussian statistics governing field quadrature measurements,
for example by homodyne detection \cite{Yue78,Tyc04}. For fermions, there is no
obvious way to extend the notion of quadratures so we omit this approach to
constructing fermion coherent states.


\section{Fermion analogy to the boson coherent state}
\label{CSfermions}

\emph{Eigenstates of the annihilation operator: ---} For fermions, the only
eigenstate of the annihilation operator with complex coefficient is the vacuum
state $|0\rangle$. Indeed, acting on a general pure state
$|\psi\rangle=\gamma|0\rangle+\delta|1\rangle$ by the annihilation operator
yields the resultant state $\delta|0\rangle$, which is not a multiple of
$|\psi\rangle$ unless $\delta=0$. However, if $|\delta|$ is small, then $\hat
c|\psi\rangle$ is close to some multiple of $|\psi\rangle$, which motivates the
consideration of ``approximate complex fermion coherent states''.

One may define  an $\epsilon$-complex fermion coherent state as follows.
For a given  $0<\epsilon<1$, an $\epsilon$-complex fermion coherent  state is a
normalized state $|u\rangle_\epsilon$, which satisfies
\begin{equation}\label{e-fermion coherent state}
  \min_{\beta\in\mathbb{C}}\bigl|\bigl| \hat{c}|u\rangle_\epsilon
        - \beta |u\rangle_\epsilon \bigr|\bigr| \le\epsilon.
\end{equation}
Here $\bigl|\bigl||\psi\rangle\bigr|\bigr|$ denotes the norm of a vector,
namely $\bigl|\bigl||\psi\rangle\bigr|\bigr|=\sqrt{\langle\psi|\psi\rangle}$.
\begin{proposition}
\label{epsilon-fermion coherent state}
For $\epsilon<1$, the normalized state
\begin{equation}
\label{appr_coh}
    |\alpha\rangle=(\cos|\alpha|+\sin|\alpha|\,\text{e}^{\text{i}\arg\alpha}\hat c^\dagger)|0\rangle
\end{equation}
is an $\epsilon$-complex fermion coherent state if $\sin^2|\alpha|\le\epsilon$.
\end{proposition}
\begin{proof}
Minimizing the norm of the vector $\hat{c}|\alpha\rangle-\beta|\alpha\rangle$
over $\beta\in\mathbb C$, we find that the minimum occurs for
$\beta=\beta_0=\frac12\sin2|\alpha|\text{e}^{\text{i}\arg\alpha}$. Evaluating
then the norm for this particular $\beta_0$, we obtain $\bigl|\bigl|
\hat{c}|\alpha\rangle-\beta_0|\alpha\rangle\bigr|\bigr|
=\sin^2|\alpha|$. Hence, if
$\sin^2|\alpha|\le\epsilon$, then $|\alpha\rangle$ is an $\epsilon$-fermion
coherent state.
\end{proof}
If $\epsilon\ll 1$, one may refer to the $\epsilon$-complex fermion
coherent state as an ``approximate complex fermion coherent state''\footnote{we
use the quotation marks to emphasize that these states, as will be seen later,
do not satisfy, even approximately, other requirements for coherent
states}. Clearly, $\epsilon$ bounds the occupation probability for a single
particle, and in the limit $\epsilon\to 0$ the set of ``approximate complex
fermion coherent states'' reduces to the vacuum state.

However, attempting to generalize such a definition of the ``approximate
complex fermion coherent state'' to a many-mode system introduces important
difficulties. Because the fermion annihilation operators of different modes are
non-commuting, it is  in principle not possible to find a common
eigenstate, with complex eigenvalue, even if a non-trivial eigenstate of
the annihilation operator of a single mode existed.  Further, to generalize
Eq.~(\ref{appr_coh}), one would need to prescribe the ordering of the actions
of the operators on the vacuum, which is a step without physical
justification. This problem will be discussed in the next paragraph.

\emph{Displaced vacuum state: ---}
It is easy to show that Eq.~(\ref{appr_coh}) can alternatively be expressed as
\begin{equation}\label{ferm_displacement}
  |\alpha\rangle=\text{e}^{\alpha\hat c^\dagger-\alpha^*\hat c}\,|0\rangle, \alpha\in\mathbb{C}.
\end{equation}
We note that $\exp(\alpha\hat c^\dagger-\alpha^*\hat c)$ is formally identical
to the boson displacement operator [see Eq.~(\ref{displaced})].  We can
therefore call $\hat D(\alpha)\equiv\exp(\alpha\hat c^\dagger-\alpha^*\hat c)$
the complex fermion displacement operator.  Any single-mode complex fermion
pure state can be obtained from the vacuum state by applying a particular $\hat
D(\alpha)$: that is, every pure fermion state is in the orbit of the vacuum
under the action of the displacement operator. If fermion coherent states were
defined as displaced vacuum states, then any single-mode fermion pure state
would be a coherent state, which  would  not yield a very useful
definition of coherent states. However, one  may  still define
``approximate complex fermion coherent states'' according to
Eq.~(\ref{ferm_displacement}) with $|\alpha|\ll1$.

To generalize this definition to a multi-mode system, one  would have to
be careful. A na\"ive approach would be to act consecutively by displacement
operators of the individual modes on the vacuum, without considering the
ordering of these operators.  However, displacement operators for different
modes neither commute nor anticommute so ordering influences the resultant
state.  For a given set of $\alpha_1,\alpha_2,\dots$, and a given permutation
$P$ of the modes $1,2,\dots$, one can define a permutation-ordered product as
\begin{align}\label{multimode-ordered}
|\vec\alpha\rangle_P \equiv
\exp\left(\alpha_{P_1}\hat{c}^\dagger_{P_1}-\alpha^*_{P_1}\hat{c}_{P_1} \right)
\exp\left(\alpha_{P_2}\hat{c}^\dagger_{P_2}-\alpha^*_{P_2}\hat{c}_{P_2} \right)
\cdots|0\rangle,
\end{align}
which depends on $P$.  For different permutations one obtains physically
different states, so one has many different multi-mode ``approximate complex
fermion coherent states'' ($N!$ variants for a finite number $N$ of occupied
modes).  This fact is demonstrated in Fig.~\ref{first-normalized}, which shows
how the complex degree of coherence (defined in Appendix~\ref{app:correlators})
depends sensitively on the choice of permutation~$P$ for two distinct multi-mode
``approximate complex fermion coherent states''.

\begin{figure}[htb]
\vspace{1mm}
(a)
\includegraphics*[width=7cm,angle=\rot,keepaspectratio]{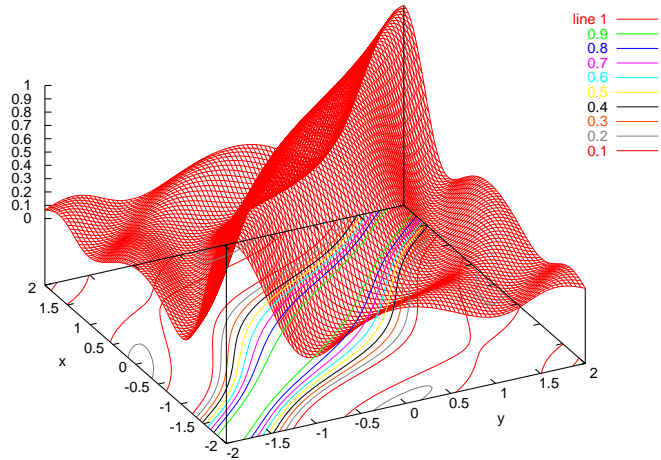}\\
(b)
\includegraphics*[width=7cm,angle=\rot,keepaspectratio]{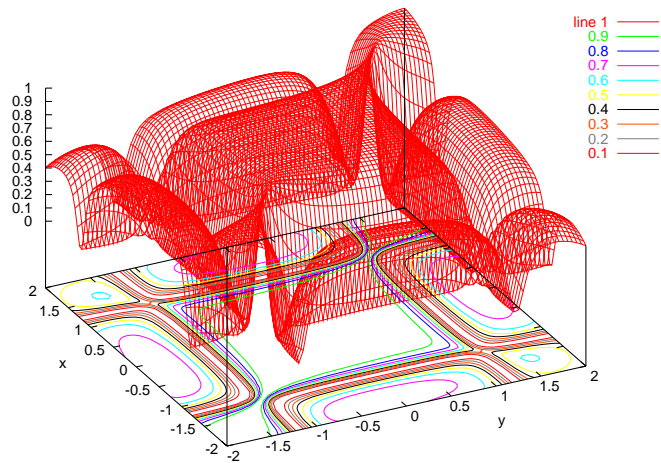}
\caption{The square of modulus of the complex degree of coherence,
$|\gamma(x,y)|^2$, for two ``approximate complex fermion coherent states''
according to Eq.~(\ref{multimode-ordered}) that differ only by the permutation
$P$. The modes are forty one-dimensional monochromatic waves with equally
spaced wavenumbers (``comb'') ordered according to their increasing wavenumber,
and all the amplitudes $\alpha_k$ are chosen to be equal to~$0.166$. The
permutation in case (a) is identity, in case (b) a random permutation. The
correlators clearly differ, which shows that the ordering of displacement
operators is of large importance and different orderings result in physically
different states.} \label{first-normalized}
\end{figure}

One could attempt to remove the permutation dependence of
Eq.~(\ref{multimode-ordered}) by averaging it over all permutations of the
modes:
\begin{equation}
\overline{ |\vec\alpha\rangle}=\sum_{P}|\vec\alpha\rangle_P,
\end{equation}
up to a normalization factor.  However, the resulting state has only zero- and
single-particle components because all the multi-particle parts cancel due to
the fermion anticommutation relations. As the state
$\overline{|\vec\alpha\rangle}$ contains no more than one fermion, it reduces
to a single-mode ``approximate complex fermion coherent state''.  Thus we
see that defining a multi-mode fermion coherent state using displacement
operators requires an a priori mode ordering. Such an ordering, however, cannot
be justified physically as none of the modes is preferred above the others.

\emph{Linear mode coupler: ---}
Suppose that the most general single-mode fermion pure state
$\beta_0|0\rangle+\beta_1|1\rangle$ is mixed with the vacuum state
via linear mode coupling. The resultant output state is
\begin{equation}\label{BS}
  \beta_0|0\rangle_1|0\rangle_2+t\beta_1|1\rangle_1|0\rangle_2
   +r\beta_1|0\rangle_1|1\rangle_2,
\end{equation}
where the indices 1, 2 label the output modes and $t$ and $r$ is the complex
transmissivity and reflectivity, respectively.  Evidently expression~(\ref{BS})
cannot be written as a product of a state of the first and second output ports,
respectively, unless $\beta_1=0$.  For small $\beta_1$, which corresponds to
the  above definition of the  ``approximate complex fermion coherent
state'', one obtains almost a product state in (\ref{BS}). Hence, ``approximate
complex fermion coherent states'' approximately satisfy the linear mode coupler
condition.  However, the condition is satisfied exactly only by the vacuum
state, that is, in the limit of no particles.

\emph{Summary: ---} Although the single-fermion ``approximate complex coherent
states'' approximately satisfy the coherence requirements given above (where,
by single-fermion, we mean a superposition of no fermion and one fermion with
the one-fermion contribution very small), we shall see that multi-mode
``approximate complex fermion coherent states'' are not so accommodating.  In
the next section we state several important propositions that hold for
normally-ordered fermion correlators and do not have direct counterparts for
boson fields. Based on these results, we will show that the correlator
factorization condition cannot be satisfied for ``approximate complex fermion
coherent states'', even approximately, except for the case of single-particle
fermion states.

\section{Properties of normally-ordered fermion correlators}
\label{FCF}

Fermion correlators are defined analogously to the boson counterparts
and provide information on expected measurement outcomes by instruments
at different spacetime coordinates. Definitions and examples
of fermion correlators are presented in Appendix~\ref{app:correlators}.

In the following we prove several important propositions that hold for
normally-ordered fermion correlators of an arbitrary state (pure or mixed) and
that are consequences of Pauli's exclusion principle.
\begin{proposition}
\label{antibunching1}
    For any fermion field state, the fermion correlator
    $\Gamma^{(n)}(x_1,\dots,x_n,y_n,\dots,y_1)$ tends to zero smoothly
    whenever two points $x_i,x_j$ or $y_i,y_j$ approach each other.
\end{proposition}
\begin{proof}
Without lost of generality we can assume that $i=1,j=2$.  Using the mode
decomposition of $\hat\psi(x_i)=\sum_{k}\varphi_k(x_i)\hat a_k$, we obtain
\begin{align}
    \Gamma^{(n)}&=\sum_{kl} \varphi^*_k(x_1) \varphi^*_l(x_{2})
     \langle\hat a^\dagger_k\hat a^\dagger_l\hat\psi^\dagger(x_3)\cdots
    \hat\psi^\dagger(x_n)\hat\psi(y_n)\cdots\hat\psi(y_1)\rangle\nonumber\\
     &=\frac12\sum_{kl}[\varphi^*_k(x_1) \varphi^*_l(x_2)-
     \varphi^*_l(x_1) \varphi^*_k(x_2)]
     \langle\hat a^\dagger_k\hat a^\dagger_l\hat\psi^\dagger(x_3)\cdots
    \hat\psi^\dagger(x_n)\hat\psi(y_n)\cdots\hat\psi(y_1)\rangle,
\end{align}
where the anticommutation relation $\hat a^\dagger_k\hat a^\dagger_l=-\hat
a^\dagger_l\hat a^\dagger_k$ was employed.

In this way, if $x_1\to x_2$, the
antisymmetric function $\varphi^*_k(x_1) \varphi^*_l(x_2)- \varphi^*_l(x_1)
\varphi^*_k(x_2)$ goes to zero and so does the whole correlator. The same
argument can be used if some pair of points $y_i,y_j$ goes together. The
smoothness follows from the smoothness of the mode functions $\varphi_k(x)$.
\end{proof}

Proposition~\ref{antibunching1} shows that it is impossible to find two
fermions at points too close to each other, which is a manifestation of
Pauli's exclusion principle.

\begin{figure}
\includegraphics[width=7cm,angle=\rot,keepaspectratio]{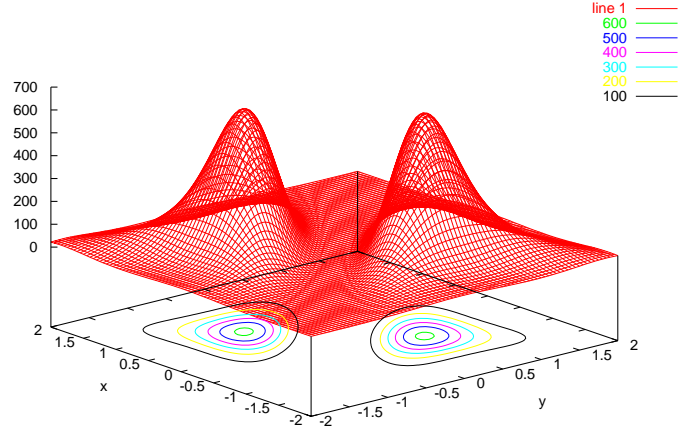}
\caption{The second-order correlator $|\Gamma^{(2)}(x,y,y,x)|^2$ for a weak
  ``approximate complex fermion coherent state'' as in
  Fig.~\ref{first-normalized}~(a).  The dip along the line $x=y$ is a common
  feature of this correlator regardless of the state and is a consequence of
  the fact that $\Gamma^{(2)}(x,y,y,x)\to0$ when $x\to y$.}
\end{figure}

\begin{proposition}\label{not-factor}
If the fermion correlator $\Gamma^{(n)}(x_1,\dots,x_n,y_n,\dots,y_1)$,
for $n>1$, factorizes for some state of the fermion field, then
$\Gamma^{(n)}(x_1,\dots,x_n,y_n,\dots,y_1)$ is identically equal to zero.
\end{proposition}
\begin{proof}
The factorization of the correlator implies that
\begin{equation}
    \Gamma^{(n)}(x_1,\dots,x_n,y_n,\dots,y_1)
        =f(x_1)\cdots f(x_n) g(y_1)\cdots g(y_n).
\end{equation}
However, if $x_i=x_j\, (i\not=j)$, then it follows from
Proposition~\ref{antibunching1} that
$\Gamma^{(n)}(x_1,\dots,x_n,y_n,\dots,y_1)=0$. Thus some of the functions
$f(x), g(x)$ must be equal zero; hence
$\Gamma^{(n)}(x_1,\dots,x_n,y_n,\dots,y_1)$ vanishes.
\end{proof}

\begin{proposition}\label{antibunching2}
Let $|\gamma(x,y)|=1$ and hence $|\Gamma(x,y)|^2=\Gamma(x,x)\Gamma(y,y)$ for
some state $|\psi\rangle$ and a pair of points $x,y$.  Then
$\Gamma^{(n)}(x,y,x_3,\dots,x_n,y_n,\dots,y_1)=0$ for any $n>1$ and any
$x_3,\dots,x_n,y_1,\dots,y_n$.
\end{proposition}
\begin{proof}
The proof is based on argument similar to the one used in \cite{Tit65,Tit66}
for bosons in the case of the electromagnetic field. Define the operator~$\hat
C$
\begin{equation}
\hat C=
 \frac1{\sqrt{\langle\hat\psi^\dagger(x)\hat\psi(x)\rangle}}\,
 \hat\psi(x)
 -\frac{\langle\hat\psi^\dagger(y)\hat\psi(x)\rangle}
  {{\langle\hat\psi^\dagger(y)\hat\psi(y)\rangle}\,
   \sqrt{\langle\hat\psi^\dagger(x)\hat\psi(x)\rangle}
}\,\hat\psi(y) .
\label{}\end{equation}
Thus $\langle\psi|\hat C^\dagger C|\psi\rangle=1-|\gamma(x,y)|^2$, which is
equal to zero by assumption.  It follows that $\hat C|\psi\rangle=0$.  Hence
\begin{equation}
 \hat\psi(x)|\psi\rangle=\frac{\langle\hat\psi^\dagger(y)\hat\psi(x)\rangle}
  {{\langle\hat\psi^\dagger(y)\hat\psi(y)\rangle}}\,\hat\psi(y)|\psi\rangle .
\end{equation}
This equation then yields $\hat\psi(x)\hat\psi(y)|\psi\rangle=0$ because
$[\hat\psi(y)]^2=0$ due to anticommutation relations.  Hence, any
normally-ordered correlator containing the operator product
$\hat\psi(x)\hat\psi(y)$ or $\hat\psi^\dagger(x)\hat\psi^\dagger(y)$ turns into
zero.  In particular $\Gamma^{(n)}(x,y,\dots)=0$.
\end{proof}

One of the consequences of Proposition~\ref{antibunching2} is that whenever the
field is perfectly first-order-coherent at the points $x,y$, it is impossible
to find a fermion at the point $x$ and another fermion at $y$.  This can be
understood as another (perhaps unusual) demonstration of the Pauli exclusion
principle.

\begin{proposition}\label{single-mode}
Let $|\gamma(x,y)|=1$ for some state $|\psi\rangle$ for all $x,y$.  Then the
state has support over only the vacuum and single-particle states.
That is, the probability for two fermions in the field is identically zero.
\end{proposition}
\begin{proof}
The proof follows directly from Proposition~\ref{antibunching2}. An alternative
proof is based on an related result for optical field stating that if
$|\gamma(x,y)|=1\,\,\forall x,y$, then only one mode of the field is
occupied~\cite{Tit66}. This holds also for fermions and in combination with
Pauli's principle it yields that there is at the most one fermion in the field.
\end{proof}

\emph{Impossibility of factorizing fermion correlators: ---}  The above
propositions, and Proposition~\ref{not-factor} in particular, show that the
fermion correlators of order larger than one cannot factorize except for the
vacuum and single-particle states.  This expresses a deep difference between
boson and fermion fields: whereas the requirement of the Glauber factorization
leads to coherent states for bosons, for fermions it leads to states 
containing zero or one particle.

Proposition~\ref{single-mode} imposes another strong condition on the fermion
field. In particular, a field with more than one fermion cannot be perfectly
first-order-coherent for all $x,y$.  Indeed, if for some field,
$\Gamma^{(1)}(x,y)=f(x)g(y)$ holds for all $x,y$, then $|\gamma^{(1)}(x,y)|=1$.
Consequently, according to Proposition~\ref{single-mode}, it follows that the
field does not contain more than one fermion. Hence, even the first-order
correlators cannot factorize for most fields. Importantly, even first-order
correlators cannot factorize for any fermion field with support over
multi-particle (even two-particle) states.

\emph{First-order correlator for ``approximate complex fermion coherent
  states'': ---} To demonstrate the result from the previous section, we will
  now calculate the first-order correlator for the multi-mode ``approximate
  complex fermion coherent state''~(\ref{multimode-ordered}) with $P$ the
  identity permutation: $P=\text{id}$.  Using the decomposition
  $\alpha_{j}=|\alpha_{j}|\text{e}^{\text{i}\phi_{j}}$ we get:
\begin{align}\nonumber
\Gamma^{(1)}(x,y)=&{}_\text{id}\langle\vec{\alpha}|\hat\psi^\dagger(x)\hat\psi(y)
                  |\vec{\alpha} \rangle_\text{id}\\
    =&\frac14 \sum_{i,j} \varphi^*_{i}(x)\varphi_{j}(y)
    \sin(2|\alpha_{i}|)\sin(2|\alpha_{j}|)\, U_{ij}\,
    \text{e}^{\text{i}(\phi_{j}-\phi_{i})} \label{1stapprox}
     +\sum_{i} \varphi^*_{i}(x)\varphi_{i}(y)\sin^4|\alpha_{i}|.
\end{align}
Here  $U_{ij}$ is the product of the factors  $\cos(2|\alpha_{k}|)$
with $k$ taking all values between $i$ and $j$ (with $i,j$ themselves
excluded). Hence, for $i+1<j$
\begin{equation}\label{Uij}
 U_{ij}=\cos(2|\alpha_{{i+1}}|)\cos(2|\alpha_{{i+2}}|)\cdots
 \cos(2|\alpha_{{j-1}}|),
\end{equation}
and for $j+1<i$ we use Eq.~(\ref{Uij}) with $i,j$ interchanged.  The factor
$U_{ij}$ comes from the anticommutation relations between the field operators.

If it were not for the factor $U_{ij}$ in Eq.~(\ref{1stapprox}), then the
correlator would approximately factorize if all $|\alpha_{i}|$ were
small. However, the factor $U_{ij}$ causes the correlator to be far from
factorization even for small $|\alpha_{i}|$. This can be seen in
Fig.~\ref{first-normalized}: if $\Gamma^{(1)}(x,y)$ factorized, then the
complex degree of coherence $|\gamma^{(1)}(x,y)|^2$ would be equal to
unity. Clearly, this does not happen in Fig.~\ref{first-normalized} although
the mean occupation number of each mode is very low:
specifically, $(\sin|\alpha|)^2=0.027$.
Comparing this situation with bosons where the correlation function
$G^{(1)}(x,y)$ factorizes  for a multi-mode coherent state,  we see that
there is a fundamental difference between fermions and bosons even in the limit
of very low occupation numbers  per mode. 

 The above results show that higher-order correlators cannot be factorized
nontrivially for any state of the fermion field and that first-order
correlators can be factorized approximately only for states containing no more
than one fermion.  In the case that there is truly zero or one particle, we
expect the features  of fermions to be the same  as those of bosons
because at least two particles are required to see manifestations of  their
quantum statistics.  However,  higher-order correlators that describe two-
or multi-particle characteristics of the field do reveal the fundamental
difference between fermions and bosons regardless of how weak the field is
because the correlator tells us about the correlation statistics regardless of
how rare the multi-particle events are. This is further demonstrated in
Appendix~\ref{Appendix:chaotic} on an example of boson and fermion chaotic
states.

Hence we see that there are no fermion states with complex parametrization that
would, at least approximately, factorize fermion correlators, with the
exception of single-particle states.  Thus the definition of coherent states in
terms of correlator factorization does not yield reasonable approximate
multi-particle fermion coherent states, similarly as the previous
definitions.

\section{Conclusion}   \label{conclusion}

Constructions of complex fermion coherent states, or approximations to such
states, is problematic, as we have shown in this paper.  We have seen that
multi-mode ``approximate coherent states'', whether obtained as displaced vacua
or as near-eigenstates of the annihilation operator, exhibit inconsistencies
due to inequivalences under permutation of modes, and also do not factorize
fermion correlation functions. The only fermion states that approximately
satisfy the coherent state definitions are the single-mode ``approximate
coherent states'' that contain, however, no more than one particle.

These problems arise ultimately from Pauli's exclusion principle and the
corresponding mathematical framework of anticommuting operators.
The ambiguity of mode ordering, the complications of linear mode coupling, and
the unfactorizability of the correlation functions are due to the exclusion
principle.  On the other hand, for states that contain no more than one
particle, the Pauli principle has no effect and ``approximate coherent states''
can be defined in the subspace of the Hilbert space spanned by zero- and
single-particle states. In this domain of the Hilbert space, there is no essential
difference between fermions and bosons as single particles cannot manifest
their distinct quantum statistics.  Hence single-mode ``approximate complex fermion coherent states''
approximately satisfy the coherence properties in the
same way as ``approximate boson coherent states'', which are obtained from
weak boson coherent states by truncating the particle number beyond one.  

It therefore appears that the use of Grassmann numbers, which obviate the
problem of anticommuting operators but introduce an unphysical parametrization,
are needed to create self-consistent fermion coherent states that satisfy the
expected coherence properties~\cite{Cah99}; complex parametrized fermion
coherent states do not appear to be useful, even approximately. 

\section*{Acknowledgements}

We appreciate valuable discussions with S.\ D.\ Bartlett, T.\ Rudolph and D.\
J.\ Rowe and acknowledge support from iCORE, CIAR, and the Australian Research
Council.

\begin{appendix}

\section{Mode ordering for mixtures of number states}
\label{chaoticSection}

A mixture of fermion number states does not suffer from the permutation
ambiguity described in Sec.~\ref{number_states}, Eqs.~(\ref{multimodef1}) and
(\ref{multimodef2}).  Indeed, the projector to a number state is invariant with
respect to changes of the mode ordering due to the doubled number of the minus
signs arising from the field operator anticommutation relations.  For example,
\begin{align}
\label{multimodec1}
 \hat P_{(1_11_2)}&=|1_1 1_2\rangle\langle1_21_1|
   =\hat a_1^\dagger\hat a_2^\dagger|0_10_2\rangle
  \langle0_20_1|\hat a_2\hat a_1 \\
 \hat P_{(1_21_1)}&=|1_2 1_1\rangle\langle1_11_2|
   =\hat a_2^\dagger\hat a_1^\dagger|0_10_2\rangle
  \langle0_20_1|\hat a_1\hat a_2 \\
 \label{multimodec2}
  \hat P_{(1_11_2)}&=\hat P_{(1_21_1)}
\end{align}
It follows that also an arbitrary mixture of number state projectors, e.g. a
chaotic state, is immune with respect to the change in ordering the modes.
This is a contrast to superpositions of number states, for which the ambiguity
is present and changing the mode ordering can lead to a physically different
state, as is  discussed in Sec.~\ref{CSfermions}.

\section{Proof of equivalence of definitions of boson coherent state}
\label{appen-equiv}

Here we prove that all the four definitions of boson coherent state mentioned
in Sec.~\ref{CSboson} are equivalent. We denote the properties as follows,\\
(a) boson coherent state is an eigenstate of the annihilation operator,\\ (b)
boson coherent state is a displaced vacuum state $|\alpha\rangle=
\text{e}^{\alpha\hat a^\dagger-\alpha^*\hat a}\,|0\rangle\equiv\hat
D(\alpha)\,|0\rangle$\\ (c) boson coherent state is a pure state that yields a
product state when mixed with the vacuum state on a linear mode coupler (linear
mode coupler). \\ (d) boson coherent state is a state for which all
normally-ordered correlators factorize.

We will show now the implications (a) $\Rightarrow$ (c), (c) $\Rightarrow$ (b),
(b) $\Rightarrow$ (a), and (a) $\Rightarrow$ (d). To show the implication (d)
$\Rightarrow$ (a) is not so easy and for the proof we refer to the paper
\cite{Tit66}.

First we will show (a) $\Rightarrow$ (c).  Let us denote the annihilation
operators of the input and output ports of the linear mode coupler by $\hat
a_1,\hat a_2$ and $\hat a_1',\hat a_2'$, respectively.  The relation between
$\hat a_1,\hat a_2$ and $\hat a_1',\hat a_2'$ is
\begin{equation} \label{linear mode couplertransform}
  \hat a_i'=\sum_{j=1}^2 A_{ij}\hat a_j
\end{equation}
with $A_{ij}$ a unitary matrix.  Let an eigenstate $|\alpha\rangle$ of $\hat
a_1$ incide on the first linear mode coupler input port, and vacuum incide on
the second input port, which is an eigenstate of $\hat a_2$ with eigenvalue
zero.  It then follows from Eq.~(\ref{linear mode couplertransform}) that the
linear mode coupler output state is an eigenstate of $\hat a_1'$ and $\hat
a_2'$, which is a product state of the linear mode coupler outputs.

Next we show (c) $\Rightarrow$ (b): Let the boson coherent state incident on
linear mode coupler be expressed in the Fock basis as follows:
\begin{equation} \label{BSincident}
  |\alpha\rangle=\sum_{n=0}^\infty c_n(\hat a^\dagger)^n|0\rangle
\end{equation}
The series $\sum_{n=0}^\infty c_n(\hat a^\dagger)^n$ formally defines a
function of the creation operator $\hat a^\dagger$ that we will denote by
$f(\hat a^\dagger)$.  The linear mode coupler input state is then $f(\hat
a_1^\dagger)|0\rangle$ as there is vacuum at the second input port.  The
relation between $\hat a_1^\dagger,\hat a_2^\dagger$ and $\hat
a_1^\dagger{}',\hat a_2^\dagger{}'$ following from Eq.~(\ref{linear mode
couplertransform}) is $\hat a_i^\dagger=\sum_{j=1}^2 A_{ji}\hat
a_j^\dagger{}'$. Therefore the output linear mode coupler state is
$f(A_{11}\hat a_1^\dagger{}'+A_{12}\hat a_2^\dagger{}')|0\rangle$. According to
our assumption, this is a product state of output modes $1',2'$, so it must
hold
\begin{equation} \label{cond_exp}
    f(A_{11}\hat a_1^\dagger{}'+A_{12}\hat a_2^\dagger{}')
  = g(\hat a_1^\dagger{}')h(\hat a_2^\dagger{}')
\end{equation}
with $g,h$ some functions.  As $\hat a_1^\dagger{}'$ and $\hat a_2^\dagger{}'$
commute, we can treat $f,g,h$ as ordinary functions of, say, arguments $x$ and
$y$ instead of $\hat a_1^\dagger$ and $\hat a_2^\dagger$.  Then, denoting
$F(z)=\ln f(z)$, making logarithm of Eq.~(\ref{cond_exp}) and performing the
derivative $\partial^2/\partial x\,\partial y$, we arrive at the condition
\begin{equation} \label{}
  A_{11}A_{12}F''(A_{11}x+A_{12}y)=0.
\end{equation}
For both $A_{11}$ and $A_{12}$ different from zero (the opposite would
correspond to a trivial linear mode coupler---either doing nothing or
interchanging the modes 1 and 2), this is possible only if $F$ is a linear
function, i.e., if $F(z)=\lambda z+\gamma$.  It then follows that $f(\hat
a^\dagger)=\text{e}^{\lambda\hat a^\dagger+\gamma}$. However, then $f(\hat
a_1^\dagger)|0\rangle$ is a displaced vacuum state (up to a normalization
constant) because the Campbell-Hausdorff-Baker formula yields
\begin{equation} \label{}
  \text{e}^{\lambda\hat a^\dagger+\gamma}|0\rangle
 =\text{e}^{|\lambda|^2/2+\gamma}\text{e}^{\lambda\hat a^\dagger-\lambda^*\hat a}|0\rangle
 =\text{e}^{|\lambda|^2/2+\gamma}\hat D(\lambda)|0\rangle.
\end{equation}

To show (b) $\Rightarrow$ (a) means to show that $\hat D(\alpha)|0\rangle$ is
an eigenstate of the annihilation operator. Displacing the annihilation
operator, we have
\begin{equation} \label{}
 \hat a\hat D(\alpha)|0\rangle=\hat D(\alpha)\hat D^\dagger(\alpha)\hat a\hat
 D(\alpha)|0\rangle=\hat D(\alpha)(\hat a+\alpha\openone)|0\rangle=\alpha\hat
 D(\alpha)|0\rangle,
\end{equation}
which we wanted to prove.

Next we show (a) $\Rightarrow$ (d): The field operators $\hat\psi(x)$ and
$\hat\psi^\dagger(x)$ are linear combinations of single-mode annihilation
operators $\hat a_i$ and creation operators $\hat a^\dagger_i$, respectively.
Any (possibly multi-mode) coherent state is therefore a right eigenstate of
$\hat\psi(x)$ and left eigenstate of $\hat\psi^\dagger(x)$. From this the
factorization~(\ref{factorization}) follows immediately.

\section{Normally-ordered correlators}
\label{app:correlators}

Correlators are important tools for describing coherence of both boson and
fermion fields. They are defined as expectation values of products of
creation and annihilation operators at different space-time points.

The fermion correlators of the $n^{\rm th}$ order are defined by
analogy to their boson counterparts
\begin{equation}
\Gamma^{(n)}(x_1,\dots,x_n,y_n,\dots,y_1)
 \equiv\langle\hat\psi^\dagger(x_1)\cdots
 \hat\psi^\dagger(x_n)\hat\psi(y_n)\cdots\hat\psi(y_1)\rangle
\end{equation}
(we use the Greek letter $\Gamma$ to distinguish from the boson
correlator~$G$).  Of particular importance is the correlator with repeated
arguments, for which we introduce the shorter expression
$\Gamma^{(n)}(x_1,\dots,x_n)\equiv \Gamma^{(n)}(x_1,\dots,x_n,x_n,\dots,x_1)$,
which corresponds to the $n$-fermion detection probability, and also the
one-fermion cross-correlator $\Gamma(x,y)\equiv\Gamma^{(1)}(x,y)$.

The normalized first-order cross-correlator
$\gamma(x,y)=\Gamma(x,y)/\sqrt{\Gamma(x,x)\Gamma(y,y)}$ is called {\em complex
degree of coherence} and its magnitude expresses the visibility of interference
fringes in a double-slit experiment in which the two slits are placed at the
points $x$ and $y$, respectively~\cite{Man95}. If $|\gamma(x,y)|=1$ for all
$x,y$, we say that the field has perfect first-order coherence.

\section{Correlators for boson and fermion chaotic states} 
\label{Appendix:chaotic}
We will show now that for chaotic states,  the fermion or boson nature of
the particles does not have an effect on the first-order correlator while it
has a strong effect on the higher-order correlators.

For a chaotic state~\cite{Gla70}, each mode of the chaotic field has maximum
entropy for a given mean number of particles in this mode, and the density
operator of the field is a direct product of single-mode density
operators. This second property simply implies that, in the chaotic state, the
individual modes of the field are totally uncorrelated.

The chaotic state density operator of a single mode of bosons or fermions with
mean number of particles $M$ can be expressed as
\begin{equation}
\hat\rho=\frac{\text{e}^{-\xi\hat a^\dagger\hat a}}{\text{Tr}\,(\text{e}^{-\xi\hat a^\dagger\hat a})},
\end{equation}
with
\begin{equation}
  \text{e}^{-\xi}=\frac{M}{1+M},\;\text{Tr}\,(\text{e}^{-\xi\hat a^\dagger\hat
  a})=1+M
\end{equation}
for bosons and
\begin{equation}
    \text{e}^{-\xi}=\frac{M}{1-M},\;\text{Tr}\,(\text{e}^{-\xi\hat
    a^\dagger\hat a})=\frac{1}{1-M}
\end{equation}
for fermions (unlike previous sections, here we denote the fermion field
operators by $\hat a,\hat a^\dagger$ rather than $\hat c,\hat c^\dagger$ for
compact notation).  Thus we have the density operators for bosons and fermions
in the following forms:
\begin{equation}
 \hat\rho_{\rm B}=\frac1{1+M}\sum_{n=0}^\infty\left(\frac
    M{1+M}\right)^n\,|n\rangle\langle n| ,\qquad \hat\rho_{\rm
    F}=(1-M)\sum_{n=0}^1\left(\frac M{1-M}\right)^n\,|n\rangle\langle n|
\end{equation}

Consider now a multi-mode chaotic field of bosons or fermions. Let there be $N$
occupied modes labeled by $1,\dots,N$ and let the mean number of particles in
the $i^\text{th}$ mode be $M_{i}$.  For brevity, we denote the boson and
fermion correlators by $G_{\rm B}$ and $G_{\rm F}$, respectively, in this
section.
The first order cross-correlator $G_\text{B,F}^{(1)}(x,y)$
can readily be calculated and has the same form for bosons and fermions:
\begin{equation} \label{chaotic-first}
  G_{\rm B,F}^{(1)}(x,y)=\sum_{i=1}^N M_{i}\varphi_{i}^*(x)\varphi_{i}(y)
\end{equation}
Eq.~(\ref{chaotic-first}) shows that the first order coherence (which is
related to the visibility of interference fringes) for chaotic states is not at
all influenced by the fermion or boson nature of the particles, provided
the mean occupation numbers are the same. This can be expected as the first
order correlator is not related to multi-particle correlations so it should not
be affected by the exchange interaction.

The $n^\text{th}$-order correlator can also be calculated for the
chaotic state. The following formula holds exactly for fermions and for bosons
it holds with good accuracy if $M_{i}\ll1, i=1,\dots,N$:
\begin{equation}\label{chaotic-nth}
  G_{\rm B,F}^{(n)}(x_1,\dots,x_n,y_n,\dots,y_1) =\sum_{P}\chi_{\rm B,F}^{{\rm
  par}(P)} G_{\rm B,F}^{(1)}(x_1,y_{P^{(1)}}) G_{\rm
  B,F}^{(1)}(x_2,y_{{P}^{(2)}})\cdots G_{\rm B,F}^{(1)}(x_n,y_{P^{(n)}})
\end{equation}
Here the sum runs over all permutations $P$ of the indices $1,2,\dots,n$, the
parity of the permutation $P$ is denoted by ${\rm par}(P)$, and $\chi_{\rm
B}=1$, $\chi_{\rm F}=-1$ is the boson and fermion sign factor, respectively.
Hence the higher-order correlators are no longer identical for boson and
fermion chaotic states, which is manifested as bunching (antibunching) for
boson (fermion) chaotic fields. If $x_i\to x_j, i\not=j$, $G_{\rm F}$ goes to
zero according to Eq.~(\ref{chaotic-nth}) due to the factor $(-1)^{{\rm
par}(P)}$, which is in agreement with the general property of fermion
correlators (Proposition~\ref{antibunching1}).

\end{appendix}


\begin{thebibliography}{99}

\bibitem {Sig05} A. I. Signal and U. Z\"{u}licke, Appl. Phys. Lett. \textbf{87}, 102102 (2005)
\bibitem {Sud63} E.~C.~G.~Sudarshan, Phys. Rev. Lett. \textbf{10}, 277 (1963).
\bibitem {Gla63a}  R.\ J.\ Glauber, Phys. Rev.  \textbf{130}, 2529 (1963).
\bibitem {Gla63b}  R.\ J.\ Glauber, Phys. Rev.  \textbf{131}, 2766 (1963).
\bibitem {Kla66} J.~R.~Klauder,  Phys. Rev. Lett. \textbf{16}, 534 (1966).
\bibitem {Mar59} J. L. Martin, Proc. Roy. Soc. Lond. A \textbf{251}, 543 (1959);
    Y. Ohnuki and T. Kashiwa, Prog. Theor. Phys. \textbf{60}, 548 (1978);
    J. R. Klauder and B.-S. Skagerstam, \emph{Coherent States}
    (World Scientific, Singapore, 1985).
\bibitem {Ber66} F. A. Berezin, \emph{The Method of Second Quantization}
                (Academic Press, New York, 1966).
\bibitem {Sch53} J. Schwinger, Phys. Rev. \textbf{92}, 1283 (1953).
\bibitem {Cah99} K.~E.~Cahill and R.~J.~Glauber, Phys.~Rev.~A~\textbf{59},
  1538 (1999).  
\bibitem {Tit65} U. M. Titulaer and R.\ J.\ Glauber, Phys. Rev.  \textbf{140},
                B676 (1965).
\bibitem {Tit66} U. M. Titulaer and R.\ J.\ Glauber, Phys. Rev.  \textbf{145},
                1041 (1966).
\bibitem {Yue78} H. P. Yuen and J.H. Shapiro, in \emph{Coherence and Quantum
                Optics IV}, edited by L. Mandel and E. Wolf (Plenum, New York,
                1978), p. 719.
\bibitem {Tyc04} T. Tyc and B. C. Sanders, J. Phys. A: Math. Gen. \textbf{37}, 7341 (2004).
\bibitem {Man95} L. Mandel and E. Wolf, Optical Coherence and Quantum
  Optics (Cambridge University Press, Cambridge, 1995).
\bibitem {Gla70} R.~Glauber, in \emph{Quantum Optics},
                edited by S. Kay and A. Maitland. (Academic Press, New York, 1970), p. 53.
                \bibitem {Aha66} Y. Aharonov, D. Falkoff, E. Lerner, and H. Pendleton,
                Ann. Phys. (N.Y.) \textbf{39}, 498 (1966).

\end{thebibliography}
\end{document}